\documentclass[aps,pre,amsmath,preprint,floatfix]{revtex4}
\usepackage{graphicx}
\usepackage{amssymb}
\begin{document}

\title{Demixing and orientational ordering in mixtures of rectangular particles}
\author{D. de las Heras}
\email{daniel.delasheras@uam.es}
\affiliation{Departamento de F\'{\i}sica Te\'orica de la Materia Condensada,
Universidad Aut\'onoma de Madrid, E-28049 Madrid, Spain}

\author{Yuri Mart\'{\i}nez-Rat\'on}
\email{yuri@math.uc3m.es}
\affiliation{Grupo Interdisciplinar de Sistemas Complejos (GISC),
Departamento de Matem\'aticas, Escuela Polit\'ecnica Superior,
Universidad Carlos III de Madrid,
Avenida de la Universidad 30, E-28911 Legan\'es, Madrid, Spain}

\author{Enrique Velasco}
\email{enrique.velasco@uam.es}
\affiliation{Departamento de F\'{\i}sica Te\'orica de la Materia Condensada
and Instituto de Ciencia de Materiales Nicol\'as Cabrera,
Universidad Aut\'onoma de Madrid, E-28049 Madrid, Spain}

\date{\today}

\begin{abstract}
Using scaled--particle theory for binary mixtures of
two--dimensional hard particles with orientational degrees of freedom, we analyse 
the stability of phases with orientational order and the demixing phase behaviour 
of a variety of mixtures. Our study is focused on cases where at least one of the 
components consists of hard rectangles, or a particular case of these, hard squares. 
A pure fluid of hard rectangles has recently been 
shown to exhibit, aside from the usual uniaxial nematic phase, an additional oriented phase, 
called tetratic phase, possessing two directors, which is the analog of the biaxial or
cubatic phases in three-dimensional fluids. There is evidence, based on computer simulation
studies, that the tetratic phase might be stable with respect to phases with lower translational 
symmetry for rectangles with low aspect ratios. As hard rectangles are mixed, in increasing 
concentration, with other particles not possessing stable tetratic order by themselves, the tetratic 
phase is destabilised, via a first-- or second--order phase transition, to uniaxial nematic or isotropic 
phases; for hard rectangles of low aspect ratio (hard squares in particular), 
tetratic order persists in a relatively large range of volume fractions. The order of these transitions 
depends on the particle geometry and dimensions, and also on the thermodynamic conditions 
of the mixture. The second component of the mixture has been chosen to be hard discs or disco--rectangles,
the geometry of which is different from that of rectangles, leading to packing 
frustration and demixing behaviour, or simply rectangles of different aspect ratio but with the same
particle area, or different particle area but with the same aspect ratio. These mixtures may be 
good candidates for observing thermodynamically stable tetratic phases in monolayers of hard particles.
Finally, demixing between fluid (isotropic--tetratic or tetratic--tetratic) phases is seen to occur in 
mixtures of hard squares of different sizes when the size ratio is sufficiently large.
\end{abstract}

   
\maketitle

\section{Introduction} 
Mixtures of three-dimensional rod--like molecules have been
analysed quite extensively, both experimentally and theoretically \cite{Lekker}.
Beautiful experiments on molecular and colloidal particles have been reported,
with the observation of different phases, such as isotropic and nematic phases of
various kinds \cite{lista}. Of special interest is the issue of orientational phase transitions 
and, in particular, of entropically--driven phase segregation in these systems,
a long--debated question in the context of hard spheres \cite{hs}.
There is now ample theoretical evidence for segregation and demixing phenomena in hard--core mixtures of 
anisotropic particles \cite{Flory,Lekker1,todos,Birsh,Vroege,Mulder0}. 
Particularly interesting are segregation phenomena between two nematic 
phases, which sometimes show upper critical points \cite{Jackson}. More recently, the occurrence and
overall influence of spatially ordered phases, in the perfect--alignment approximation,
have been analysed \cite{Koda,Mulder0}, and more complete studies, lifting the latter approximation,
on the effect of layered phases and microsegregation phenomena in the phase behaviour 
have been carried out \cite{Giorgio1,Giorgio2,Giorgio3}.

Studies on the corresponding two--dimensional mixtures are very 
scarce. In Ref. \cite{Talbot} it was shown that two--dimensional isotropic mixtures 
of hard convex bodies can never demix within scaled--particle theory (SPT). A unimodal 
polydisperse mixture of hard needles was also studied within the Onsager approach in
Ref. \cite{Schoot}, one result being that the isotropic--nematic transition
is always of second order. Also, an equation of state for two--dimensional mixtures of 
hard bodies has been constructed starting from an approximation for their direct correlation 
functions \cite{Perera_0}; it was found that demixing never occurs. Finally, a theoretical study 
has recently been carried out for 
mixtures of hard rectangles and discorectangles within the framework of SPT \cite{Martinez-Raton1}.
Using a bifurcation analysis, demixing between different phases, one of which is an 
orientationally--ordered phase, was shown to occur.

The analysis of this problem is sufficiently 
motivated by the importance of surface--phase transitions experienced
by monolayers of adsorbed molecules. But, since the transition from the
disordered phase to the oriented (nematic) phase in one--component fluids
may be, and in most cases is, of second order in two dimensions, many new interesting 
features may arise in the phase diagram of the mixture, such as tricritical
and critical end points, which are absent in three dimensions. 
Our recent work on two--dimensional hard--rod fluids \cite{Martinez-Raton1}
has demonstrated that these systems do in fact exhibit a richer phase behaviour, 
with the additional fact that, due to spatial restrictions, phase behaviour may be
more sensitive to subtle effects associated with the particle 
geometrical shapes. It would be desirable to understand the differences 
and similarities between the two dimensionalities and elucidate their origin.

In fact, liquid-crystalline phase transitions in one--component
fluids depend strongly on dimensionality. A general
trend of two--dimensional systems with continuous symmetry 
is the lack of true long--range order
\cite{Strandburg},
which in a two--dimensional nematic would be reflected in the presence 
of quasi--long--range orientational order \cite{Straley,Tobochnik,Frenkel,Bates,Lagomarsino}. 
Thus, in the absence of any 
other mechanism, transitions from the isotropic phase (I) 
to the uniaxial nematic phase (N$_{\hbox{\tiny u}}$) in two--dimensional, one--component fluids 
of hard rods, may be governed by a disclination unbinding--type
mechanism \cite{Frenkel,Bates,Lagomarsino}, as proposed by the Kosterlitz--Thouless (KT) theory \cite{Kosterlitz}
(in fluids of hard ellipses, for which simulations exist \cite{Frenkel,VB,Cuesta}, the nature 
of the transition seems to depend on the aspect ratio \cite{Cuesta}). Mean--field theories,
which cannot account for these effects since collective fluctuations and the dynamics
of topological defects are not properly (or not at all) described, predict continuous transitions of
the usual (mean--field) type \cite{Kayser,Cuesta-Tejero-Baus,Schlacken}. 
However, the direct role played by dimensionality in two--dimensional mixtures
may be of secondary importance
as far as entropically-driven {\it demixing} transitions are concerned, since local
entropic effects 
associated with packing may completely preempt KT--type effects;
in this sense, the mechanisms governing these systems
could be intimately connected with those operating in the corresponding three--dimensional
mixtures, which are described qualitatively correctly by mean--field theories
of the density--functional type \cite{Lekker}.

Of particular interest is the case of hard rectangular 
particles, which might exhibit a so--called tetratic phase (N$_{\hbox{\tiny t}}$)
in the one--component fluid 
\cite{Schlacken,Martinez-Raton2,Wojciechowski,Donev,Three-body}. 
The tetratic phase possesses two equivalent directors pointing 
along mutually orthogonal directions; it exhibits a symmetry higher than that of the 
particles making up the fluid. The nature of the transition from the I to the
N$_{\hbox{\tiny t}}$ phases is unknown, but density--functional studies predict
it to be of second order \cite{Schlacken,Martinez-Raton1}. A recent bifurcation analysis
\cite{Martinez-Raton2},
combined with further calculations which include three-body correlations \cite{Three-body}, have not 
been conclusive as to the absolute thermodynamic stability of the tetratic phase
with respect to phases with spatial order,
although preliminary computer simulations \cite{Wojciechowski,Donev,Three-body} seem to support the tetratic 
phase as an intermediate phase between the isotropic phase 
and the crystalline phase, at least for aspect ratios 
less than $\sim 7$--$9$ \cite{Three-body}. 
However, questions remain as to the maximum aspect ratio that can support 
tetratic order (for example, experiments on -non--equilibrium-- vibrated monolayers of granular 
particles \cite{Narayan} extend it up to $\sim 12$), the interplay between
the N$_{\hbox{\tiny t}}$ and the usual uniaxial nematic phase N$_{\hbox{\tiny u}}$
(possessing only one director or alternatively two non--equivalent directors), 
or the role played by the nature of the crystalline tetratic phase (with an aperiodic solid
being the most promising candidate \cite{Donev}). Finally, recent work \cite{MR2} has 
shown the ability of these particles to promote spatial order when 
confined between parallel one--dimensional plates, with potentially 
interesting applications as building blocks for self--assembly.

The role played by tetratic order in mixtures where one of the components
consists of hard squares or rectangles, while the other may or may not promote 
this order, is the aim of the present paper. We investigate the phase 
behaviour of mixtures of hard squares or hard rectangles with other particles of
different geometry. The ability of hard rectangles
to induce either short-- or long--range tetratic order originates from the
sharp corners of their shape (here we do not address the question on the existence
of true long--range order of nematic correlations in these systems). Rectangles with low aspect ratio $\kappa=L/\sigma$
(with $L$ their length and $\sigma$ their breadth) may orient along either
a particular direction or the orthogonal direction equally easily without
hampering packing efficiency, and a tetratic phase results. For larger aspect ratios,
hard rectangles stabilise into the usual uniaxial nematic. Hard-rod particles 
terminated by a semicircle (such as a disco--rectangle) do not pack efficiently
in a tetratic arrangement, and can only form one (uniaxial) nematic phase. When
these two types of particles, or when two types of hard rectangles, one
exhibiting a tetratic phase which is not present in the other, mix together,
the tetratic order induced by one component survives, at higher pressures, in
some range of particle volume (area) fractions, with an extent that depends on the 
geometrical compatibility between species.
Nematic demixing occurs, and the resulting phase diagrams show a rich variety of 
features, such as phase transitions of different order, triple, critical, 
tricritical, azeotropic and critical end points. 

A particular situation is when hard squares of different particle area are mixed. 
A consensus is now beginning to emerge that entropy-driven demixing in additive hard--sphere 
mixtures involves spatially non--uniform phases \cite{hs}. The current situation concerning
mixtures of {\it parallel} (i.e. with frozen orientational degrees of freedom) hard cubes 
looks very similar \cite{3D}. Now, despite initial evidence
based on simulation that no demixing occurs in a mixture of parallel hard squares on a lattice
\cite{Dijkstra}, supported by off--lattice density--functional calculations on the fluid phase
\cite{Cuesta2}, recent simulation work gives evidence for demixing involving
an inhomogeneous phase \cite{Buhot}. In contrast, our calculations, which incorporate
orientational degrees of freedom (though, admittedtly, do not contemplate inhomogeneous
phases), show demixing involving two fluid phases, at least one of which is oriented
(with an island of instability in the phase diagram and sometimes
with an associated upper critical point), when the size ratio is sufficient large,
while the completely isotropic mixture does not exhibit segregation. Clearly
our results demand for additional computer simulations on the freely--rotating hard--square model,
which have not been reported yet.

After presenting a brief summary of the theoretical approach, which is a 
density--functional theory based on the SPT approximation, we present
the results and conclude with a summary and some final remarks. The Appendix 
presents some details on the calculation of spinodal lines and tricritical points.

\section{Theory} 

Let us first briefly recall the theoretical model used and the approximations
implemented. The model is based on the SPT approximation for two-dimensional 
binary mixtures, first applied in three--dimensions by 
Cotter and Wacker \cite{Cotter}. The first extension to the two--dimensional case,
a model of hard rectangles with restricted orientations \cite{Scott}, was later
rederived for isotropic mixtures of general hard convex bodies \cite{Talbot}.
Here we use the implementation of the SPT approximation for oriented mixtures derived, 
using the standard procedures, in Ref. \cite{Martinez-Raton1}, to which the reader 
is referred for further details.
Let us denote the free--energy density $f$ in units of the thermal energy $kT$ by $\Phi$, i.e.
$\Phi=f/kT$. The density functional for the corresponding excess (over ideal gas) quantity,
$\Phi_{\hbox{\tiny exc}}[h_1,h_2]$, depends on the two orientational distribution 
functions $h_{\nu}(\phi)$ ($\nu=1,2$) for the two components of the mixture, $\phi$ being
the angle between the particle main axis and some reference direction in the plane, 
which is arbitrarily taken as the $x$ axis. In SPT 
approximation is written as \cite{Martinez-Raton1}
\begin{equation}
\Phi_{\hbox{\tiny exc}}[h_1,h_2]=\rho\left\{-\log{\left(1-\eta\right)}+
\frac{\rho}{2\left(1-\eta\right)}\sum_{\nu\tau}x_{\nu}x_{\tau}
\left<\left<V_{\nu\tau}^{(0)}\right>\right>\right\}.
\label{SPT}
\end{equation}
Here the subindeces $\nu,\tau=1,2$ refer to the two components of the mixture, 
$\rho=\rho_1+\rho_2$ is the total density, with $\rho_{\nu}$
the density of species $\nu$, $\eta =\rho_1v_1+\rho_2v_2$ is the total packing fraction, 
with $v_{\nu}$ the particle area for species $\nu$, and the number fractions are defined as 
usual by $x_{\nu}=\rho_{\nu}/\rho$. Also, $V_{\nu\tau}^{(0)}(\phi)=
V_{\nu\tau}(\phi)-v_{\nu}-v_{\tau}$, where $V_{\nu\tau}(\phi)$ is the angle--dependent 
excluded volume between species $\nu$ and $\tau$. The double angular average
$\left<\left<V_{\nu\tau}^{(0)}\right>\right>$ is defined by
\begin{equation}
\left<\left<V_{\nu\tau}^{(0)}\right>\right>=
\int_0^{2\pi}d\phi\int_0^{2\pi}d\phi^{\prime}h_{\nu}(\phi)
V_{\nu\tau}^{(0)}(\phi-\phi^{\prime})h_{\tau}(\phi^{\prime}).
\end{equation}
The functions $V_{\nu\tau}(\phi)$ are analytical; their expressions 
were explicitely written in Ref. \cite{Martinez-Raton1}, except that 
corresponding to the cross interaction between rectangles and discorectangles
\cite{CrossHRHDR}.

Note that the one--component limit of Eqn. (\ref{SPT}) correctly reduces to 
the Onsager theory (where only the second virial coefficient
is incorporated) as the density vanishes, since SPT recovers the exact
second virial coefficient; however, in contrast to the three--dimensional case, the Onsager
theory is not rigorously correct in the hard-needle limit in two dimensions,
since three-- and higher--order virial coefficients are {\it not} vanishingly
small in this limit, some of them being even negative \cite{Rigby}).
Therefore, higher--order contributions in density are only approximately reproduced by 
Eqn. (\ref{SPT}). SPT can be considered
as a sophisticated Onsager theory, in that spatial (not orientational, which are still included 
up to second order in density) correlations are somehow resumed into density--dependent terms.
Alternative approaches, such as a two--dimensional version of the Parsons--Lee approach
for three--dimensional hard rods, have the same structure and, for lack of a detailed
performance analysis, can in principle be considered to be equivalent. The ideal contribution
\begin{equation}
\Phi_{\hbox{\tiny id}}[h_1,h_2]=\sum_{\nu}\rho_{\nu}\left(\log{\rho_{\nu}}-1+
\int_0^{2\pi} d\phi h_{\nu}(\phi)\log{\left[2\pi h_{\nu}(\phi)\right]}\right),
\end{equation}
is added to obtain the complete free energy-functional $\Phi[h_1,h_2]=
\Phi_{\hbox{\tiny id}}[h_1,h_2]+\Phi_{\hbox{\tiny exc}}[h_1,h_2]$.
Functional minimisation of $\Phi[h_1,h_2]$ with respect to $h_{\nu}(\phi)$ gives the
equilibrium configuration of the mixture. The pressure follows from the equation
\begin{equation}
\frac{p}{kT}=\frac{\rho}{1-\eta}+\frac{\rho^2}{2(1-\eta)^2}
\sum_{\nu\tau}x_{\nu}x_{\tau}\left<\left<V_{\nu\tau}^{(0)}\right>\right>.
\end{equation}

Instead of numerically solving the Euler--Lagrange equations associated with
$\Phi[h_1,h_2]$, we follow 
common practice in our research group and tackle the direct minimisation of the functional.
Various strategies are possible \cite{Strategies}. In the present article we choose to introduce a
parameterized form for the orientational distribution functions, $h_{\nu}(\phi)$.
To facilitate computations, these functions are simply parameterised as 
\begin{equation}
h_{\nu}(\phi)=\frac{\displaystyle e^{\Lambda_{\nu}^{(1)}\cos{2\phi}+\Lambda_{\nu}^{(2)}\cos{4\phi}}}
{\displaystyle\int_0^{2\pi} d\phi^{\prime}
e^{\Lambda_{\nu}^{(1)}\cos{2\phi^{\prime}}+\Lambda_{\nu}^{(2)}\cos{4\phi^{\prime}}}}
\label{Parameter}
\end{equation}
The parameters $\Lambda_{\nu}^{(k)}$ take care of the two types of orientational 
symmetries, either uniaxial ($k=1$) or tetratic ($k=2$). Equivalently two
order parameters, $q_{\nu}^{(k)}$, can be defined as
\begin{equation}
q_{\nu}^{(k)}=\int_0^{2\pi} d\phi h_{\nu}(\phi) \cos{\left(2k\phi\right)},\hspace{0.4cm}
k=1,2,
\end{equation}
which are proportional to the coefficients of a Fourier-expansion of the functions $h_{\nu}(\phi)$
including the two lowest symmetries that a rectangular particle can generate.
Table \ref{tabla} summarises the different phases with their associated values of 
the $\Lambda_{\nu}^{(k)}$ and $q_{\nu}^{(k)}$ parameters in a one--component phase.
\begin{table}
\begin{tabular}{c|r|r|r|r}
\hline\hline
\hbox{\bf phase} & $\Lambda^{(1)}$ & $\Lambda^{(2)}$ &  $q^{(1)}$ & $q^{(2)}$\\
\hline\hline
\hbox{isotropic, I} & $0$ & $0$ & $0$ & $0$ \\
\hline
\hbox{uniaxial nematic, N$_{\hbox{\tiny u}}$} & $> 0$ & $\ge 0$ & $>0$ & $>0$\\
\hline
\hbox{tetratic nematic, N$_{\hbox{\tiny t}}$} & $0$ & $>0$ & $0$ & $>0$\\
\hline\hline
\end{tabular}
\caption{Possible values of the variational parameters $\Lambda^{(k)}$ and
order parameters $q^{(k)}$ in the isotropic (I), uniaxial nematic (N$_{\hbox{\tiny u}}$) and
tetratic (N$_{\hbox{\tiny t}}$) phases.}
\label{tabla}
\end{table}
The equilibrium configurations of the mixtures are more conveniently obtained by minimising the
Gibbs free energy per particle $g=(p+f)/\rho$ with respect to the $\Lambda_{\nu}^{(k)}$ 
parameters at fixed value of the pressure $p$ and composition $x\equiv x_1$
(we will henceforth arbitrarily associate $x$ with the number fraction of the component
labelled as 1). Minimisations were performed using an efficient routine based on 
the Newton--Raphson method. Coexistence (binodal) lines were located by means of 
a standard common--tangent construction. 

The use of the parameterisation (\ref{Parameter}) obviously introduces an
approximation over the exact calculation, and it would be of interest to 
know the amount of error introduced. As far as the calculation of bifurcation or
spinodal lines of the various phase transitions is concerned, the parameterisation has 
no impact, since first--order terms in $\Lambda_{\nu}^{(k)}$, 
\begin{equation}
h_{\nu}(\phi)=\frac{1}{2\pi}\left(1+\Lambda_{\nu}^{(1)}
\cos{2\phi}+\Lambda_{\nu}^{(2)}\cos{4\phi}\right)+
O\left(\Lambda_{\nu}\right)^2,
\end{equation}
which appear as quadratic terms in the free energy, are treated exactly.
However, binodal lines and tricritical points are affected by the
parameterisation, since their location depend on higher--order terms in $\Lambda^{(k)}$.
We have checked the parameterisation in two ways: first, some selected
calculations have been performed using an additional cosine term, $\cos{6\phi}$ ($\cos{8\phi}$), in the
parameterisation for the uniaxial (tetratic) nematic phase; the differences found, at the level of coexistence lines,
were found immaterial. Second, tricritical points have been evaluated
exactly by computing the exact fourth--order terms (involving the above cosine terms depending 
on the symmetry of the phase). In all cases the
differences with respect to the calculations using (\ref{Parameter}) have 
been found to be of minor importance. The analysis is presented
in the Appendix, which contains quantitative details on this issue.

\section{Results} 
Results are presented, in the form of pressure--composition phase diagrams, in
Figs. \ref{fig1}--\ref{fig8}. 
We divide the presentation by first showing results for different
mixtures containing squares and
discs, followed by mixtures of hard rods (rectangles and discorectangles).
In the following we will generally use the following acronyms for the
different particles: HS (hard squares), 
HD (hard discs), HR (hard rectangles) and HDR (hard discorectangles).

\subsection{Mixtures of hard squares}
\label{Sub1}

\begin{figure}
\mbox{\includegraphics*[width=3.5in, angle=0]{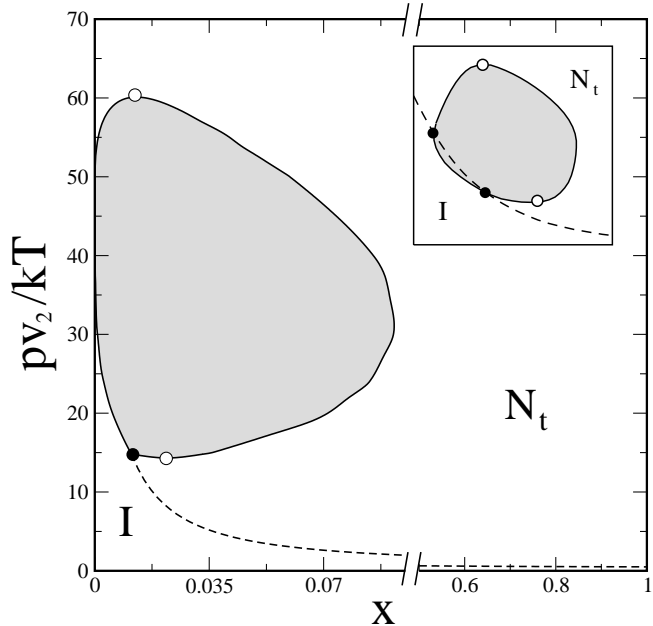}}
\caption{Phase diagram of a hard-square ($L_{\nu}=\sigma_{\nu}$ for $\nu=1,2$)
mixture with $\kappa_1=10$ and $\kappa_2=1$ 
in the scaled--pressure $pv_2/kT$ vs. composition $x=x_1$
plane. Open circles: critical points. Filled circles: critical end points.
Inset is a qualitative scheme depicting the relative locations
of the I--N$_{\hbox{\tiny t}}$ spinodal and the region of demixing.
Two-phase region is indicated by the gray area. Note that the scale of the
composition (horizontal) axis is discontinuous.}
\label{fig1}
\end{figure}

The fluid of freely--rotating hard squares has been studied by Monte Carlo simulations
\cite{Wojciechowski}, with indications that a tetratic phase
might be stable prior to crystallisation. On the other hand, mixtures of
hard squares have been analysed by computer simulations \cite{Dijkstra,Buhot} and 
density--functional theory \cite{Cuesta2} in the approximation of perfect orientational
order. Indications that there is no demixing in this system \cite{Dijkstra,Cuesta} were later 
challenged by evidence for a spinodal line from Monte Carlo simulation \cite{Buhot};
the demixing transition involves a fluid phase and a phase with big squares in 
close--packed aggregates. Since orientational disorder may be important for these
mixtures, we believe it is of interest to address this problem with the SPT approximation,
where orientational disorder is allowed (obviously our approach can only
search for demixing behaviour involving fluid, either isotropic or tetratic, phases).

The results shown in Fig. \ref{fig1}
correspond to a mixture of freely--rotating squares with size ratio 1:10, and
have been gathered in a pressure--composition phase diagram, with label 1 
assigned to the larger particles. According to SPT theory, the fluids of the unmixed 
(one--component) species both exhibit corresponding second--order I--N$_{\hbox{\tiny t}}$ phase transitions 
at reduced pressure $pv/kT\simeq 0.52$ (here $v$ is the particle proper area). 
Note that, for squares, a uniaxial nematic phase is not possible
by construction. For the mixture, the only possible (fluid) phases are also the I and 
N$_{\hbox{\tiny t}}$ phases. Remarkably, 
the mixture exhibits segregation between two tetratic phases with different area fractions
occupied by the two components. The region of demixing is a closed
loop bounded above (below) by an upper (lower) critical point (in fact, since the 
I--N$_{\hbox{\tiny t}}$ spinodal line crosses the demixing region --see
inset showing schematic topology--, segregation mostly proceeds
between I and N$_{\hbox{\tiny t}}$ phases). Since mixtures of perfectly parallel 
hard squares do not demix it is surprising that, in this mixture, orientational disorder 
induces demixing.

Closed loops of immiscibility have been predicted in mixtures of hard rods in the
Onsager approximation \cite{Birsh,Vroege} (see however Ref. \cite{Mulder}), 
and also in fluids of parallel hard rods 
\cite{DuBois}, both in three dimensions. Our calculations show that they are also
a property of some two--dimensional mixtures. In our case demixing occurs 
at fairly low values of composition; this is easy to explain since we
expect segregation to take place when the volume fraction of both
components are approximately equal, $\eta_1\simeq\eta_2$, which implies that
segregation will occur for $x\sim v_2/v_1=10^{-2}$. As the size ratio of the
squares decreases, the island of immiscibility diminishes, and we are eventually
left with a second--order I--N$_{\hbox{\tiny t}}$ transition in the whole composition 
interval. The demixing island disappears when the size ratio is equal to $1:4$
(see Appendix). 

The mechanism underlying N$_{\hbox{\tiny t}}$--N$_{\hbox{\tiny t}}$ demixing
is different for the upper and lower parts of the demixing island.
As is well known, demixing and ordering phenomena in mixtures of hard anisotropic particles
result from the competition of entropies of different origin: mixing,
orientational and excluded--volume entropies. In the transition from the isotropic
to the nematic phase the last two terms compete. In mixtures, the excluded--volume entropy
has contributions from the two species and from the unlike--particle interactions.
The mechanism explaining the lower N$_{\hbox{\tiny t}}$--N$_{\hbox{\tiny t}}$ segregation 
phenomena in Fig. \ref{fig1} is the classical one: the balance between mixing entropy and 
the excluded--volume term of unlike species, which counterbalances the tendency toward mixing 
of the former. However, the upper N$_{\hbox{\tiny t}}$--N$_{\hbox{\tiny t}}$ segregation region 
has a different origin, since here orientational entropy plays a role. This might have been 
suspected a priori, since in the limit of perfect order (parallel squares) no demixing occurs. 
At sufficiently high pressures in the freely--rotating fluid, sufficiently near the
close--packed limit, the orientational order is almost saturated and can no longer compete with 
the other terms, so that no demixing is expected. As pressure is reduced, orientational
entropy begins to play a role and in fact the contribution from the small squares is the
driving force toward demixing. 

For the size ratio shown (1:10), two critical end points, indicated in the figure by
filled circles, appear in the phase diagram, defined by the points where the
I--N$_{\hbox{\tiny t}}$ spinodal line crosses the demixing region; the location of
these points, given by $x^{(1)*}=0.012$, $p^{(1)*}v_2/kT=13.86$ and 
$x^{(2)*}=8\times 10^{-5}$, $p^{(2)*}v_2/kT=49.38$, can be approximated from the set of
candidates to tricritical points obtained by means of a bifurcation analysis 
(here and in what follows, numerical values for the location 
of tricritical and points should be understood to result from rigorous bifurcation analysis 
of the free--energy functional, and may in some cases be at variance with those obtained from the
variational minimisation, on which all the phase diagrams presented are based; 
see Appendix for details on the bifurcation analysis). As the size ratio
is diminished, first the upper critical point disappears, and the upper critical end 
point becomes a tricritical point; then the lower critical end point becomes a
tricritical point, and the lower critical point also disappears, leaving two
tricritical points before the whole demixing region vanishes. The lower critical point always stays 
within the N$_{\hbox{\tiny t}}$ region (i.e. no I--I demixing is observed), .

Demixing in this mixture is therefore associated with a symmetry--breaking phase, the
tetratic phase. There are strong arguments \cite{Talbot} disallowing 
demixing in the I phase (within the context of the SPT approach).
This would not necessarily imply that demixing is completely ruled out in the 
rotationally--symmetric I phase of hard--square mixtures or mixtures of particles with
different geometries, but a more sophisticated theory, incorporating exact higher--order
virial coefficients (which are known to be important for two--dimensional hard convex 
bodies), would be necessary to settle this point.

\begin{figure}
\mbox{\includegraphics*[width=3.5in, angle=0]{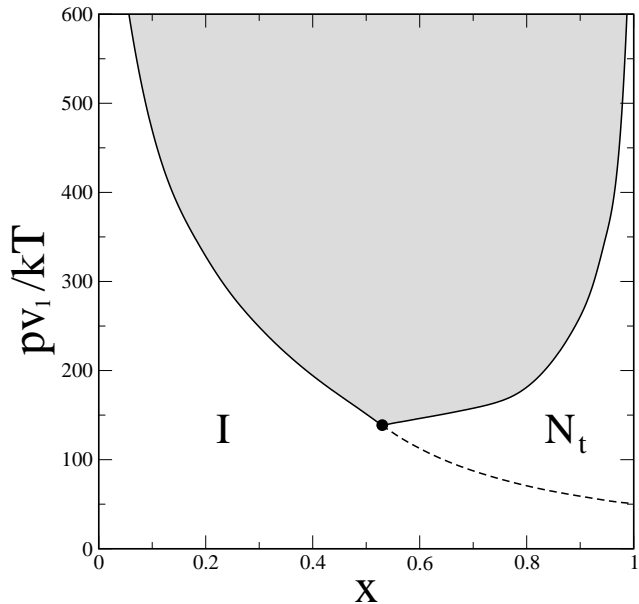}}
\caption{Phase diagram for a HS/HD mixture in the scaled--pressure $pv_1/kT$ vs. 
composition $x=x_1$ plane, with $v_1=L_1\sigma_1$ the volume of the squares. 
The side length of the squares, $L_1=\sigma_1$, is chosen to be 
the same as the diameter of the discs, $\sigma_2$.
Filled circle indicates tricritical point.
Two-phase region is indicated by the gray area.}
\label{fig2}
\end{figure}

\subsection{Mixtures containing squares and discs}

Fig. \ref{fig2} shows the phase diagram corresponding to a mixture of hard discs
and squares (the diameter of the former being equal to the side length 
of the latter). Here only one of the components (the squares) has tetratic order. 
By choosing the particle areas of both species to be approximately equal
(the ratio being $\simeq 0.79$) we focus on the role of particle geometry. 
In this mixture we find that the N$_{\hbox{\tiny t}}$ phase exhibited by the pure system of squares
survives up to a maximum concentration of discs of $\sim 50\%$ (this corresponds to
an area ratio approximately equal to $0.79$). Complete demixing occurs 
at high pressure, while the I--N$_{\hbox{\tiny t}}$ transition becomes of second order
below a tricritical point, located at $x=0.57$, $pv_1/kT=131.95$ (see Appendix).
Fractionation becomes stronger as pressure is
increased, since excluded--volume considerations are increasingly important in this limit:
geometrical mismatch (given by the unlike disc-square interaction) grossly counterbalances
mixing entropy, inducing strong segregation.

\begin{figure}
\mbox{\includegraphics*[width=3.5in, angle=0]{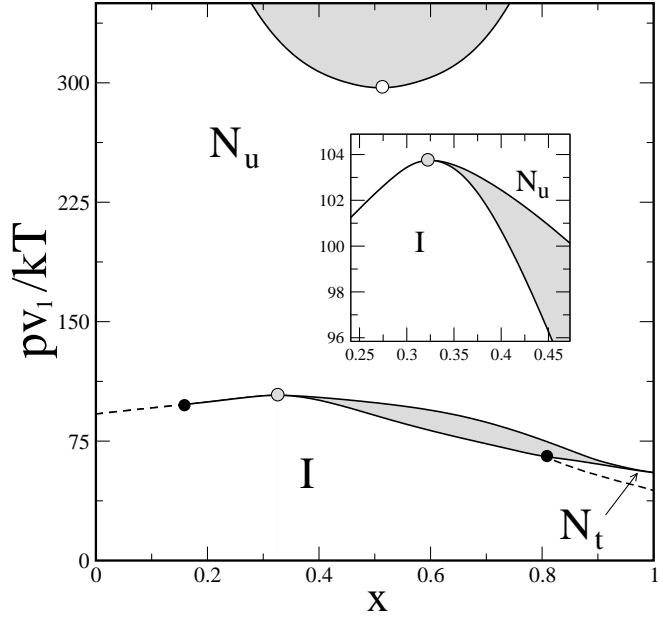}}
\caption{Phase diagram for a HR/HDR mixture in the scaled--pressure $pv_1/kT$ vs.
composition $x$ plane. The two components have the same aspect ratio ($\kappa_1=L_1/\sigma_1=2$
for the rectangles and $\kappa_2=(L_2+\sigma_2)/\sigma_2=2$ for the discorectangles),
and same particle area. Open circle indicates critical point, shaded circle denotes an azeotropic 
point, while filled circle indicates tricritical point.
Two-phase regions are indicated by gray areas.}
\label{fig3}
\end{figure}


\subsection{Mixtures of rectangles and discorectangles}

Hard rectangles have been predicted to exhibit a phase with tetratic order when their
aspect ratio is 
sufficiently low \cite{Schlacken,Martinez-Raton2,Wojciechowski,Donev,Three-body}.
Tetratic order is enhanced as the aspect ratio of 
the rectangles is decreased. However there is some
uncertainty as to the critical value of $\kappa$ beyond which the tetratic phase
is no longer possible; the SPT approach indicates that for $\kappa<2.21$ the
stable nematic phase is the tetratic nematic \cite{caution}
(as mentioned in the introduction, 
computer simulations indicate that this value may be much larger, in the
range $\sim 7$--$9$). In contrast, a fluid of hard discorectangles 
can only support a uniaxial nematic phase \cite{Bates}.

In previous work \cite{Martinez-Raton1} we analysed possible demixing scenarios
of mixtures of hard rectangles (HR) and mixtures of hard discorectangles (HDR),
using the same SPT theory. In this section we further investigate this problem
by considering a wider range of mixtures, in particular, {\it crossed} mixtures
(i.e. mixtures consisting of HR and HDR particles); the effect of particle
geometry is an aspect that can be assessed in a very direct way by analysing
crossed mixtures, as well as the role played by the tetratic phase against the
standard uniaxial nematic phase when particles of different geometries are mixed.
Results will be presented by means of phase diagrams including,
not only spinodal lines, but also binodal lines when present.

In the following we first consider mixtures of hard rods, one of which can be stabilised
into a tetratic phase (i.e. HR particles with aspect ratio $\kappa<2.21$) 
while the other cannot (i.e. either HR particles with $\kappa>2.21$, or HDR particles). 
We intend to understand how the tetratic phase is destabilised by the
geometrical mismatch of the particles.

The first mixture that we consider is
a mixture of HR and HDR particles, both with $\kappa=2$ [in the latter case 
the aspect ratio parameter is defined as $\kappa=(L+\sigma)/\sigma$]. 
Also, the same particle areas 
have been chosen for both types of particles, in an attempt to single out
features of phase behaviour mainly driven by differences in particle geometry. The 
resulting phase diagram is depicted in Fig. \ref{fig3}. At high pressure a demixing 
region occurs between two uniaxial nematic phases (i.e. there is 
N$_{\hbox{\tiny u}}$--N$_{\hbox{\tiny u}}$ demixing), bounded by a lower critical point.
Fractionation becomes stronger as pressure increases, an effect ultimately
associated with the slight difference in particle geometry, which penalises
the mixed state due to unfavourable packing between dissimilar particles; this
arises from the circular caps of the HDR particles. At lower pressure
there is a transition between the isotropic and the uniaxial nematic phases, 
which occurs via a first--order phase transition with small fractionation. 
The phase diagram is of the azeotropic type: an azeotropic point 
(no fractionation) appears at $x\simeq 0.32$, with an associated (small) pressure range 
where the N$_{\hbox{\tiny u}}$ phase is reentrant.
%
Since the HR component of the mixture possesses a stable tetratic phase, an island of tetratic order exists
for high values of $x$, separated from the isotropic phase via a second--order
phase transition and from the uniaxial nematic via a first--order transition
(of course this is also the case for the pure HR fluid, $x=1$). The island
of N$_{\hbox{\tiny t}}$ stability is very small; for a HR fluid with $\kappa=2$,
the range of pressures where the tetratic phase is stable is small \cite{Martinez-Raton2} but,
in addition, tetratic order is easily destroyed when adding to the mixture particles 
that do not conform with tetratic symmetry. 

\begin{figure}
\mbox{\includegraphics*[width=3.5in, angle=0]{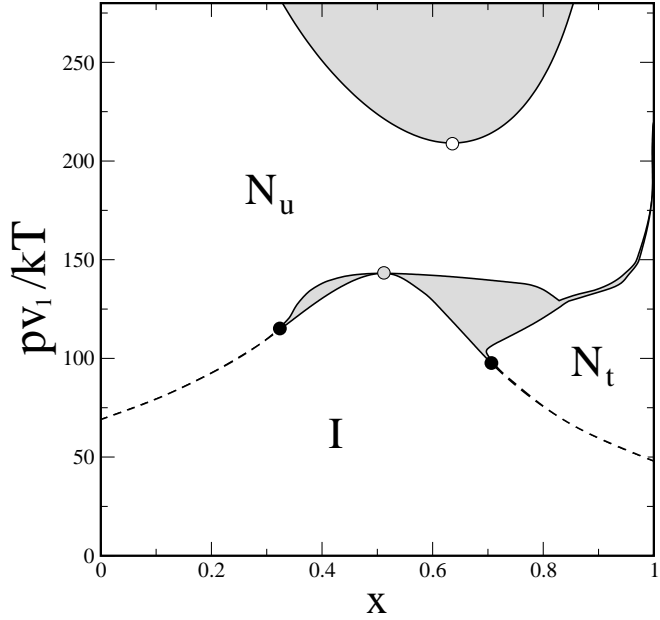}}
\caption{Phase diagram for a HR/HDR mixture in the scaled--pressure $pv_1/kT$ vs.
composition $x$ plane. Values of the parameters are $\kappa_1=1.5$, $\sigma_1=1$
for the rectangles and $\kappa_2=2$ and same particle area as a rectangle of 
aspect ratio equal to 2 and unit breadth for the discorectangles.
Open circle indicates critical point, shaded circle denotes an azeotropic point,
while filled circles indicate tricritical points.
Two-phase regions are indicated by the gray areas.}
\label{fig4}
\end{figure}

Next we consider two different mixtures that represent slight variations with
respect to the previous mixture. In Fig. \ref{fig4} we analyse a HR/HDR mixture,
where the HR component has been shortened to an aspect ratio of $\kappa_1=1.5$,
keeping the breadth to the same value, while the HDR component still has 
$\kappa_2=2$ but the particle area is made equal to that of the HR particle
of the previous mixture. Essentially we would like to analyse the effect of
shortening the hard--rectangle component in the previous mixture (while
at the same time reducing its particle area by $25\%$) with the aim of 
increasing the strength of tetratic ordering.
Since the tetratic phase of the one--component HR fluid is much more stable,
the tetratic phase in the mixture stabilises into a larger range of compositions. The
diagram is topologically equivalent to that in Fig. \ref{fig3}, except
that the critical end point in the I--N$_{\hbox{\tiny t}}$ spinodal line
now becomes a tricritical point, and there appears a (new)
I--N$_{\hbox{\tiny u}}$--N$_{\hbox{\tiny t}}$ triple point.

\begin{figure}
\mbox{\includegraphics*[width=3.5in, angle=0]{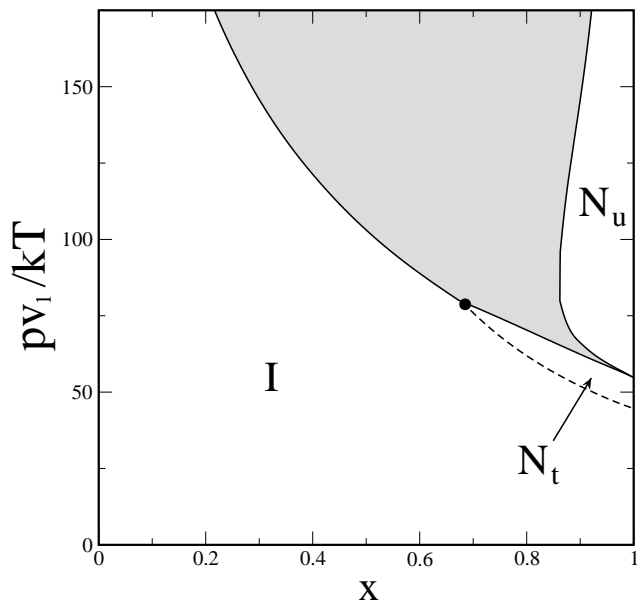}}
\caption{Phase diagram for a HR/HDR mixture in the scaled--pressure $pv_1/kT$ vs.
composition $x$ plane. Values of the parameters are $\kappa_1=2$, $\sigma_1=1$ for
the rectangles and $\kappa_2=1.5$ and same particle area as rectangle of aspect
ratio equal to 1.5 and unit breadth for the discorectangles. Filled circle
indicates critical end points. Two-phase region is indicated by the gray area.}
\label{fig5}
\end{figure}

Non--trivial changes are obtained if the mixture of Fig. \ref{fig3} is
changed by shortening the HDR particle down to an aspect ratio of $\kappa_2=1.5$
while the particle area is made equal to that of a HR particle of aspect
ratio 1.5 and unit breadth; as a result, the nematic phase of the
one--component HDR fluid appears at a much higher pressure. The phase diagram is 
presented in Fig. \ref{fig5}. This phase diagram can be regarded as a continuation 
of that in Fig. \ref{fig3} where the N$_{\hbox{\tiny u}}$--N$_{\hbox{\tiny u}}$ 
demixing region and the I--N$_{\hbox{\tiny u}}$ coexistence have collapsed into 
a single demixing region. The I--N$_{\hbox{\tiny t}}$ phase stability region now increases,
as a result of the HDR particle having a lower proper area. 

\begin{figure}
\mbox{\includegraphics*[width=3.5in, angle=0]{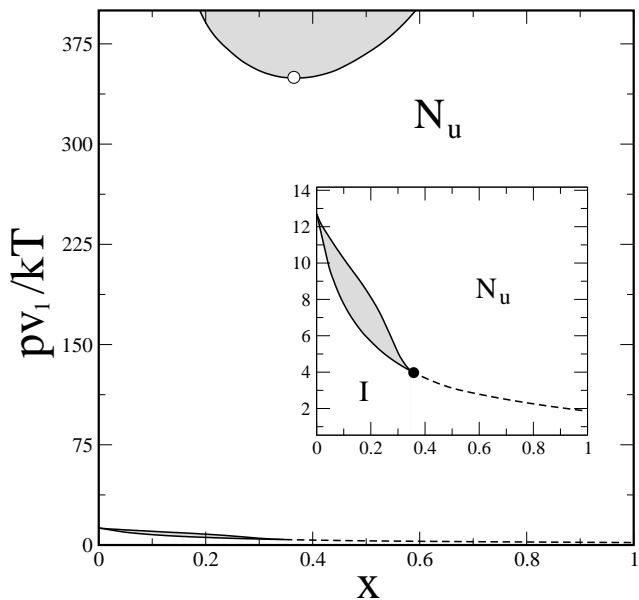}}
\caption{Phase diagram for a HR/HR mixture in the scaled--pressure $pv_1/kT$ vs.
composition $x$ plane. Particle breadths are set to the same value, while
aspect ratios are $\kappa_1=10$ and $\kappa_2=5$. Open circle: critical point.
Filled circle: tricritical point. Two-phase regions are indicated by the gray areas.}
\label{fig6}
\end{figure}

\subsection{Mixtures of rectangles}

\begin{figure}
\mbox{\includegraphics*[width=3.5in, angle=0]{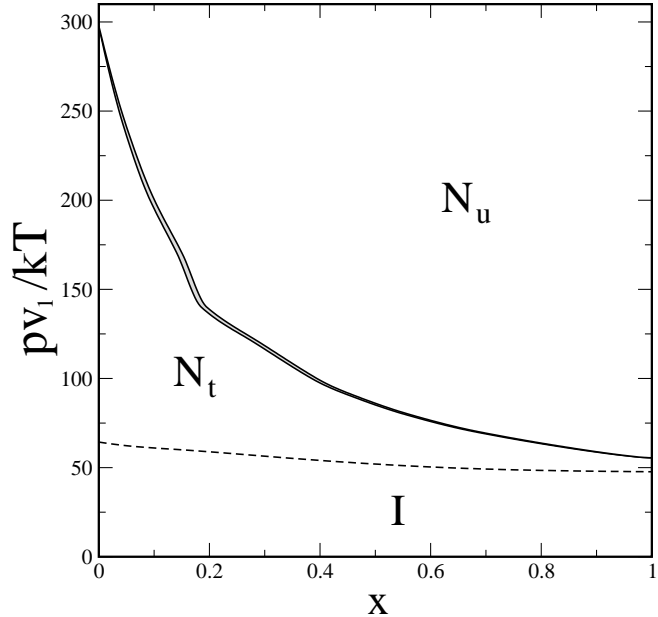}}
\caption{Phase diagram for a HR/HR mixture in the scaled--pressure $pv_1/kT$ vs.
composition $x$ plane. Particles have the same breadth, with $\kappa_1=2$ and 
$\kappa_2=1.5$. Two-phase region is indicated by the gray area.}
\label{fig7}
\end{figure}

\begin{figure}
\mbox{\includegraphics*[width=3.5in, angle=0]{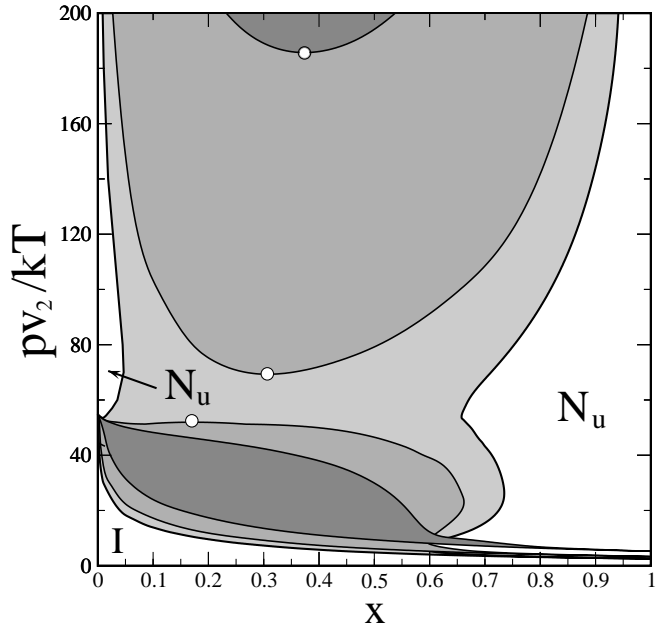}}
\caption{Phase diagram for a HR/HR mixture in the scaled--pressure $pv_2/kT$ vs.
composition $x$ plane. All rectangles are taken to have the same breadth,
and $\kappa_2=2$. The first component has $\kappa_1=4.0, 4.6$ and $5.0$, 
corresponding to two-phase regions in lighter gray in the graph.}
\label{fig8}
\end{figure}

Finally we consider mixtures of HR particles. Here the breadth of all particles
will be taken to be unity, and we change the length or, equivalently, the
aspect ratio. Three cases are considered: (i) $\kappa_1=5$, $\kappa_2=10$,
so that none of the components exhibits tetratic symmetry; (ii) 
$\kappa_1=2$, $\kappa_2=1.5$, with both species having tetratic phases;
and (iii) $\kappa_1$ in the range $4.0$--$5.0$, $\kappa_2=2$, so that
only the second species can stabilise into a tetratic phase.
This cases are shown in Figs. \ref{fig6}-\ref{fig8}, respectively.

In mixture (i) no tetratic phase appears (Fig. \ref{fig6}). There is 
N$_{\hbox{\tiny u}}$--N$_{\hbox{\tiny u}}$ 
demixing at high pressure. The I--N$_{\hbox{\tiny u}}$ transition 
is of first order, but becomes of second order at a
tricritical point (see inset), located at $x=0.35$, $pv_1/kT=4.07$ (see Appendix).

The aspect ratios of the two components of mixture (ii), $\kappa_1=2$ and
$\kappa_2=1.5$, were chosen such that
both possess a stable tetratic phase. This requires their values to be very
similar; therefore, no demixing occurs (Fig. \ref{fig7}), 
and a relatively featureless phase diagram results.

Finally, in mixture (iii), the shorter component exhibits a tetratic
phase that propagates from the $x=0$ axis into finite values of composition
when the longer component is added; however, due to the considerable difference in 
lengths between the
two species, the N$_{\hbox{\tiny t}}$ phase does not survive very much as a 
function of composition; in fact, the region of tetratic stability can
hardly be seen in the phase diagram, Fig. \ref{fig8}. 
At high pressure there is a large demixing region which,
as the particle aspect ratio of the first component is increased
(from $4.0$ to $5.0$), expands considerably.
Again demixing is due to unfavourable excluded--volume interactions between
unlike species. In general, the phase diagram varies considerably even
for very slight variations in particle shape of the first species.
For $\kappa_1=4.0$ and $4.6$, a separate region 
associated with the first--order I--N$_{\hbox{\tiny u}}$ transition 
appears at low pressures; in some cases (cf. the case $\kappa_1=4.6$)
reentrant behaviour in the nematic phase (stronger than in three--dimensional 
hard--rod mixtures --see e.g. Ref. \cite{Lekker1}), is found. Also, for the case
$\kappa_1=4.6$ an {\it upper} critical point associated with
N$_{\hbox{\tiny u}}$--N$_{\hbox{\tiny u}}$ demixing can be seen. This feature
has been seen in three--dimensional mixtures of thin and thick hard rods of
the same length \cite{Varga}; in our two--dimensional mixture the particle thickness 
is the same, and it is particle length that is different.
Therefore, in some range of parameters two N$_{\hbox{\tiny u}}$--N$_{\hbox{\tiny u}}$ 
coexistence regions can be found, with upper and lower critical points. These
two regions merge as $\kappa_1$ increases, giving rise to a very large 
demixing region.

\section{Discussion and concluding remarks}

To conclude, we have analysed the phase behaviour of mixtures of two-dimensional
hard rods, using the scaled--particle theory approximation. We have particularly 
concentrated on mixtures containing hard 
rectangles as one of the components, with the other being a hard discorectangle, what we
have called {\it crossed} mixtures. The geometry of hard rectangles allows for a tetratic
phase to be stabilised; addition of a second component which does not support 
this symmetry
obviously tends to destroy tetratic order but, when the particle areas of both components
are not very dissimilar and the pure fluid of hard rectangles strongly stabilises the
tetratic phase (i.e. when their aspect ratio is relatively low), tetratic order may survive even for
large area fractions of the non--tetratic--forming component.
The case of a mixture of hard rectangles, one with tetratic order and the other without,
shows the same trends. 

As a special case we have also considered mixtures of hard squares, and found that these
mixtures do exhibit nematic demixing, contrary to the case where rotational degrees of freedom
are frozen; in the latter case the mixture never segregates into two fluid phases
(though simulations seem to point to segregation between fluid and non-uniform phases), 
while in the former there is a critical size ratio (1:4) above which there is segregation. 
This is a remarkable case where orientational {\it disorder} induces entropy--driven demixing (i.e.
{\it mixing order}). Inclusion of non-uniform phases in our theoretical analysis would 
certainly be interesting and is left for future work. Growing evidence indicates that 
demixing in additive hard--core mixtures always involves at least one inhomogenous phase. 
An interesting possibility is that freely--rotating hard--core mixtures (which can be
viewed as a special case of non--additive mixture) may in some cases segregate into two 
fluid phases, provided nematic phases are involved; it could be that some type of
symmetry breaking, either positional or orientational, is required for demixing to occur.

In general, phase diagrams of two--dimensional hard--rod particles closely
resemble those of three-dimensional mixtures, save the fact that the isotropic--nematic
transition in two dimensions may be (and actually is in most cases) of second order, which adds
an element of complexity to the corresponding phase diagrams. Nematic--nematic demixing generally occurs
in these systems and, for sufficiently large size ratios between the components,
demixing competes with the isotropic--nematic (either uniaxial or tetratic)
transition. Entropic effects due to balance between excluded-volume interactions
of like and unlike components and mixing entropy produce strong fractionation. These effects 
probably counteract collective fluctuations leading to KT-type behaviour;
for example, within our mean--field density--functional treatment, the
continuous isotropic--nematic transition of the one--component fluid generally
continues as a critical transition in the mixture, but sooner or later,
beyond some value of particle composition, a demixing (first--order) transition 
is met via a tricritical point or otherwise (critical end point).
However, the true nature of the isotropic--nematic transition in these mixtures,
and of the uniaxial and tetratic nematic phases themselves (i.e. whether they possess 
true or quasi--long range order) is an issue that should be investigated by means
of detailed computer simulations and possibly also by experiments on vibrated monolayers.

In this work we have not considered spatially--ordered phases, such as
smectic, columnar or crystalline. Certainly some of these phases will appear 
at some pressure, and probably some of the phase behaviour shown for
the mixtures considered will be preempted by these phases. More
sophisticated density--functional treatments are required to assess this point;
some proposals have already been done for one--component two--dimensional
hard--rod systems \cite{Martinez-Raton2}, but even the consequences of these approaches 
have not been explored yet; this will be the subject of future work \cite{delasHeras}.

\begin{acknowledgments}
Y.M.-R. was supported by a Ram\'on y Cajal research contract. 
This work is part of the research
projects MOSAICO, FIS2005-05243-C02-01 and FIS2004-05035-C03-02
of the Ministerio de Educaci\'on y Ciencia (Spain), and S-0505/ESP-0299 of
Comunidad Aut\'onoma de Madrid (Spain).
\end{acknowledgments}

\begin{appendix}
\section{Bifurcation analysis}

In Ref. \cite{Martinez-Raton1} two of us carried out a bifurcation analysis to study
the nature of phase transitions in two--dimensional hard--rod fluid mixtures and
to calculate the location of the tricritical points present in their phase diagrams.
We refer the reader to this work for details on the calculations. For the sake of
completeness, a brief summary of the main ingredients of the bifurcation analysis is 
presented here.

The Fourier series representation of the orientational distribution functions
of the two different species, labelled as $\nu=1,2$, is
\begin{eqnarray}
h_{\nu}(\phi)=\frac{1}{2\pi}\left[1+\sum_{k\ge 1}^{\infty}h_k^{(\nu)}\cos(2k\phi)\right].
\end{eqnarray}
After inserting these expressions in the free-energy per particle $\varphi=\Phi/\rho$ and expanding
in Taylor series with respect to the Fourier amplitudes $h_k^{(\nu)}$, we obtain
a Landau expansion for $\Delta\varphi=\varphi_N-\varphi_I$, the free--energy difference between 
the orientationally ordered and isotropic phases, in terms of these amplitudes.
Further, minimizing $\Delta\varphi\left(\{h_k^{(\nu)}\}\right)$ with respect to all
amplitudes except one (here chosen as $h_i\equiv h_i^{(1)}$, with $i=1,2$ for uniaxial and
tetratic nematic phases, respectively), and substituting the result back
in $\Delta\varphi$, keeping terms up to fourth order, we obtain the expression 
\begin{eqnarray}
\Delta\varphi=Ah_i^2+Bh_i^4,
\end{eqnarray}
The coefficients $A(x,\eta)$ and $B(x,\eta)$ are both functions of the
composition $x\equiv x_1$ and packing fraction $\eta$. The spinodal curve
of the transition between the isotropic and nematic (uniaxial or tetratic) phases
can be calculated as $A(x,\eta^*)=0$, which defines the packing fraction
$\eta^*(x)$ as a function of the composition.

It can be shown \cite{Martinez-Raton1} 
that the instability region of the mixture with respect to composition and
volume fluctuations defines a region in the plane $(\kappa_1,x)$, at fixed $\kappa_2$,
bounded by curves calculated as the roots of the function $T_N^*$, defined by
\begin{eqnarray}
T_N^*=T_I^*-\frac{1}{2B^*}\left[
\frac{\partial}{\partial y}\left(y^2\frac{\partial \varphi_I}
{\partial y}\right)^*\left(A_x^*\right)^2+
\left(\frac{\partial^2\varphi_I}{\partial x^2}\right)^*
\left(y^*A_y^*\right)^2-2\left(\frac{\partial^2\varphi_I}
{\partial x\partial y}\right)^*\left(y^*\right)^2
A_x^*A_y^*\right],\nonumber\\
\end{eqnarray}
where
\begin{eqnarray}
T_I=\frac{\partial}{\partial y}
 \left(y^2\frac{\partial \varphi_I}{\partial y}\right)
\left(\frac{\partial^2 \varphi_I}{\partial x^2}\right)
-\left(y\frac{\partial^2\varphi_I}{\partial x\partial y}\right)^2.
\end{eqnarray}
Here $y=\rho/(1-\eta)$, and the asterisk over any function of $y$ means
that this function is to be evaluated at $y^*$, its bifurcation value (note
the different definition of $y$ with respect to the one used in Ref. 
\cite{Martinez-Raton1}; both coincide when $v_{\nu}=1$, the constraint used in Ref. 
\cite{Martinez-Raton1}). Finally, $A_x^*$ and $A_y^*$ are the partial derivatives of $A$
with respect to $x$ and $y$, respectively, both evaluated at $y^*$.

\subsection{Mixtures of hard squares}
The packing fraction at the spinodal line obtained as the solution
of $A(x,\eta^*)=0$ is a constant, independent of $x$,
and is equal to $\eta^*=\left[1+8/(15\pi)\right]^{-1}$.
Further, the I-N$_t$ tricritical points can be calculated as the roots
of
\begin{eqnarray}
T_{N_t}^*&=&\frac{16}{x_1x_2}\left\{
1-\frac{2}{11}s_3^2\left[1+s_2^2+15\frac{(s_2-s_1)^2}
{1+s_1^2}\right]\right\},
\end{eqnarray}
with
\begin{eqnarray}
s_i=\sqrt{\frac{\langle\sigma^{2i}\rangle}
{\langle\sigma^i\rangle^2}-1},\quad i=1,2,\hspace{0.6cm}
s_3=\sqrt{\frac{11\langle \sigma^2\rangle^3}
{16\langle\sigma^4\rangle\langle\sigma^2\rangle-
5\langle\sigma^3\rangle^2}},
\end{eqnarray}
where we defined $\langle u^n\rangle=\sum_i x_iu_i^n$.
Introducing the new variable $\xi=x_2v_2/\langle v\rangle$, and after some
lengthy but straightforward calculations, we arrive at
\begin{eqnarray}
T_{N_t^*}=\frac{16}{x_1x_2}\frac{\left(25z^2+(34-16r)z+9\right)}
{\left(11+z(16r+6-5z)\right)},
\label{lo}
\end{eqnarray}
where the new variable $z=(r-1)\xi$ was introduced, and we defined
$r=\sigma_1/\sigma_2$. The quadratic equation of $z$
given in the numerator of (\ref{lo}) has the following roots:
\begin{eqnarray}
\xi_{1,2}=\frac{1}{25(r-1)}\left[8r-17\pm4\sqrt{
(r-4)(4r-1)}\right].
\label{sol}
\end{eqnarray}
We can see from this expression
that demixing can only occur for $r > 4$. The pressure
at the tricritical point can be calculated using the value of
the roots found above, resulting in
\begin{eqnarray}
\frac{p^*\sigma_1^2}{kT}=\frac{15\pi}{8r^2}
\left[r^2-(r^2-1)\xi+\frac{15}{2}\left[r-(r-1)\xi\right]^2\right].
\end{eqnarray}
For $r=10$ we find
\begin{eqnarray}
\xi_{12}=\frac{1}{75}\left(21\pm4\sqrt{26}\right),
\end{eqnarray}
with the approximate values $\xi_1\simeq 0.5519$ and
$\xi_2\simeq 0.0081$. Using
$x_2=\xi/\left[r^2-(r^2-1)\xi\right]$, we find $x_2^{(1)*}\simeq 0.0122$
and $x_2^{(2)*}\simeq 8.12\times10^{-5}$, while the values for
the pressures are
$p^{(1)*}\sigma_1^2/kT\simeq 13.8603$ and
$p^{(2)*}\sigma_1^2/kT\simeq 49.3842$.

\begin{figure}
\mbox{\includegraphics*[width=3.5in, angle=0]{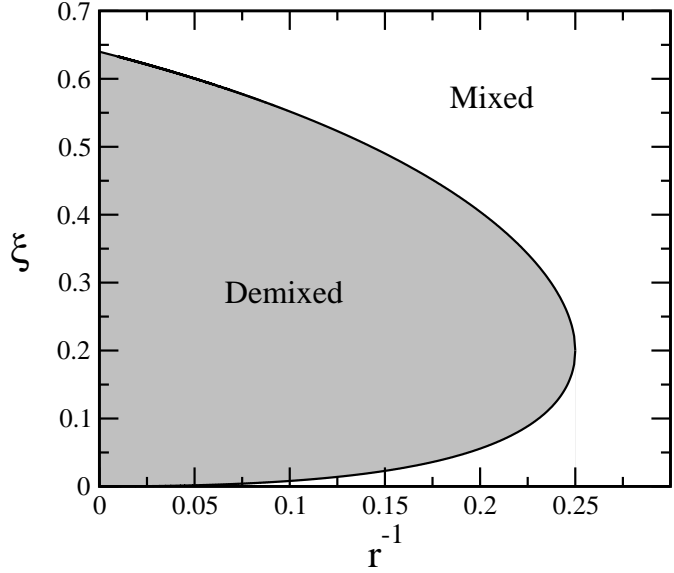}}
\caption{The mixed and demixed states that follow from the solution to
$T^*_{N_t}=0$, with $T^*_{N_t}$ given by Eqn. (\ref{lo}).}
\label{App1}
\end{figure}

In Fig. \ref{App1} are shown the mixed and demixed states of
the hard--square mixture in the $\xi$--$r^{-1}$ plane, as obtained from the
roots in Eqn. (\ref{sol}).
We have checked that this is indeed the scenario for
values of $r$ not too much larger than its critical value $r^*=4$, i.e.
the I-N$_t$ transition changes from second (first) to first (second)
order above (below) the lower (upper) tricritical point.
As we have already shown, the
phase diagram for $r=10$ has two critical points, but the
tricritical points are located close to the
critical end points (see Fig. \ref{fig1}).

\subsection{Mixtures of hard squares and hard discs}

The solution to the equation $A(x,\eta^*)=0$ gives us the following expression
for the isotropic-tetratic spinodal curve:
\begin{eqnarray}
\eta^*=\left[1+\frac{8x\sigma_1^2}{15\pi \langle v\rangle}\right]^{-1},
\end{eqnarray}
while the solution to $T_{N_t}^*=0$ can be found also analytically as
\begin{eqnarray}
x^*=\left[1+\frac{16}{35}\left(\frac{4\sigma_1}{\pi\sigma_2}\right)^2\right]^{-1}.
\label{ala}
\end{eqnarray}
Finally, the pressure at the tricritical point can be calculated as
\begin{eqnarray}
\frac{p^*\sigma_1^2}{kT}=\frac{15\pi}{8x^*}\left\{
1+\frac{15}{2x^*\sigma_1^2}\left[\frac{\pi\sigma_2}{4}+
\left(\sigma_1-\frac{\pi\sigma_2}{4}\right)x^*\right]^2\right\}.
\end{eqnarray}
For $\sigma_1=\sigma_2=1$ we find
$\left(x^*,p^*v_1/kT\right)\simeq(0.5744,131.9459)$. The value for the composition 
calculated from the minimization is about $0.53$ (see Fig. \ref{fig2}). This difference is due to the
parametrization used. One should include, in the exponential parameterization of
the orientational distribution functions, terms such as
$\Lambda_{\nu}^{(2)}\cos 4\phi+\Lambda_{\nu}^{(4)}\cos 8\phi$
to properly take into account the tetratic symmetry about the tricritical point.

\subsection{Mixtures of hard rectangles}

For mixtures of hard rectangles the isotropic--uniaxial nematic
spinodal calculated as the solution to $A(x,\eta^*)=0$ gives
\begin{eqnarray}
\eta^*=\left[1+\frac{2}{3\pi}\frac{\langle(L-\sigma)^2\rangle}{
\langle v\rangle}\right]^{-1}.
\end{eqnarray}
The value of the pressure at bifurcation is
\begin{eqnarray}
\frac{p^*v_1}{kT}=y^*\left[1+\frac{3}{2}\frac{\langle(L+\sigma)\rangle^2
}{\langle(L-\sigma)^2\rangle}\right].
\end{eqnarray}
Solving $T_N(x,\eta^*)=0$ numerically with respect to $x$
for a particular mixture with $(L_1=10,\sigma_1=1)$ and $(L_2=5,\sigma_2=1)$,
we find a value of $x^*\simeq 0.3472$, which defines the location of the
tricritical point. The packing fraction and the pressure at this
point are $\eta^*\simeq 0.4515$ and $p^*v_1/kT\simeq 4.0659$,
respectively.

\end{appendix}


\begin{references}
\bibitem{3D} The situation in three--dimensions (parallel hard cubes) is not
completely settled. Initial simulations on mixtures of parallel hard cubes
on a lattice \cite{Dijkstra}, supported by an off--lattice Fundamental--measure theory (FMT)
study \cite{Cuesta2}, seemed to indicate evidence for demixing between two fluid phases.
A more detailed analysis of a more consistent off-lattice FMT functional \cite{YuriCuesta}, and more recently 
a FMT--based formulation on a lattice \cite{Lafuente}, conclude that demixing involves
one spatially--ordered phase, and that fluid--fluid demixing is metastable. The present
state of affairs is therefore similar to that in hard--sphere mixtures.
\bibitem{CrossHRHDR} The excluded volume between a hard rectangle (labelled as 1) and a hard 
discorectangle (labelled as 2) is $V_{12}(\phi)=\sigma_1\sigma_2+L_1\sigma_2+L_2\sigma_1
|\cos\phi|+L_1L_2|\sin\phi|+v_1+v_2$, with $v_1=L_1\sigma_1$ and $v_2=L_2\sigma_2+\pi\sigma_2^2/4$ the 
areas of the rectangle and the discorectangle, respectively.
\bibitem{Strategies} For the implementation of a Fourier expansion, see Ref.
\cite{Martinez-Raton2}; parameterised forms were used in the present work and
also in Refs. \cite{Martinez-Raton1,Three-body}.
\end{references}
\end{document}